\documentclass[]{jkas} 
\usepackage{subcaption} 
\usepackage{multirow}
\usepackage[normalem]{ulem}

\def\year{2023} 
\def\volume{56} 
\def\issue{1} 
\def\beginpage{1} 
\def\received{January 31, 2026} 
\def\accepted{XXXXXX XX, 2026} 
\def\published{XXXXXX XX, 2026} 
\date{Received \received; Accepted \accepted; Published \published}

\usepackage{siunitx}
\usepackage{subcaption}
\usepackage{color}

\title{%
SQUIDPOL: Seoul National University QUadruple Imaging Device for POLarimetry
}

\author[1]{Sunho Jin\thanks{Co-first author}}{0000-0002-0460-7550}
\author[2]{Jooyeon Geem\thanks{Co-first author}}{0000-0002-3291-4056}
\author[3, 4]{Masateru Ishiguro}{0000-0002-7332-2479}
\author[1]{Woojin Park}{0000-0001-8012-5871}
\author[1]{Heeyoung Oh}{0000-0002-0418-5335}
\author[1]{Chan Park}{}
\author[3, 4]{Seungwon Choi}{0009-0006-1028-1653}
\author[1]{Yoonsoo P. Bach}{0000-0002-2618-1124}
\author[5]{Hyeonwoo Ju}{0009-0001-8488-1786}
\author[3, 4]{Jinguk Seo}{}
\author[3, 4]{Bumhoo Lim}{0000-0002-8244-4603}
\author[3, 4]{Myungshin Im}{0000-0002-8537-6714}

\affil[1]{ Korea Astronomy and Space Science Institute (KASI), 776 Daedeok-daero, Yuseong-gu, Daejeon 34055, Republic of Korea}
\affil[2]{Asteroid Engineering Laboratory, Lule\aa\ University of Technology, Box 848, SE-98128 Kiruna, Sweden}
\affil[3]{Department of Physics and Astronomy, Seoul National University, 1 Gwanak-gu, Seoul 08826, Republic of Korea}
\affil[4]{SNU Astronomy Research Center, Department of Physics and Astronomy, Seoul National University, 1 Gwanak-gu, Seoul 08826, Republic of Korea}
\affil[5]{Department of Physics, Konkuk University, 120 Neungdong-ro, Gwangjin-gu, Seoul 05029, Korea}

\def\corrauthor{%
Masateru Ishiguro, \email{ishiguro@snu.ac.kr}
}

\def\runningauthor{%
Jin and Geem et al.
}

\def\runningtitle{%
SQUIDPOL: Seoul National University QUadraple Imaging Device for POLarimetry
}

\def\keywords{%
journals: individual: JKAS
}

\def\abstracttext{
We present SQUIDPOL, a low-cost, multi-channel optical imaging polarimeter that performs simultaneous linear polarization measurements using a rotating half-wave plate, a non-polarizing beam splitter, and four wire-grid filters.
We show that the off-the-shelf non-polarizing beam splitter introduces measurable polarization-dependent systematics, which can bias polarimetric measurements if left uncorrected. We quantify this effect for both transmitted and reflected beams and incorporate a correction scheme into the data-analysis pipeline.
On-sky validation demonstrates stable and reproducible performance, achieving a polarization accuracy of $\sigma_\mathrm{P} \sim$ 0.15\% for bright polarized standard stars. Mounted on the 60-cm Ritchey-Chr\'etien telescope (the focal length of 4200 mm, f/7) at the Pyeonchang Observatory of Seoul National University, SQUIDPOL provides an effective common field of view of $13.5' \times 8.2'$ with a pixel scale of $0.45"$ pixel$^{-1}$ and supports standard $B$, $V$, $R_\mathrm{C}$, and $I_\mathrm{C}$ filters.
}

\begin{document}
\jkashead 

\section{Introduction\label{sec:intro}}

Polarimetry has made significant contributions to observational astronomy, offering unique insights into the physical properties of astronomical targets and their surrounding environments. For example, measurements of stellar polarization enable investigations of interstellar magnetic fields via the extinction and alignment of dust grains \citep{2015ARA&A..53..501A}. Synchrotron radiation, which exhibits strong polarization as a result of the relativistic motion of electrons in magnetic fields, serves as a powerful diagnostic tool for magnetized environments, such as supernova remnants and active galactic nuclei \citep[e.g.,][]{1980ARA&A..18..321A}. In solar system science, polarimetry also plays a crucial role. Light reflected from airless bodies (e.g., asteroids and satellites) exhibits linear polarization that depends on the surface properties such as geometric albedo and particle size, allowing researchers to infer the physical characteristics of their surfaces \citep[e.g.,][]{2024A&A...684A..80B,2024A&A...688A.195G}. Moreover, light scattered by atmospheres or aerosols of planets in the solar and extrasolar systems is polarized, which can be used to study atmospheric composition, particle distribution, and radiative transfer processes \citep{2000ApJ...540..504S}. Given these diverse applications, polarimetry remains a fundamental observational technique in both astrophysical and planetary research, providing critical information that cannot be obtained through photometry or spectroscopy alone.


To meet the aforementioned scientific requirements for astronomical polarimetry, observatories worldwide have developed and operated dedicated polarimetric instruments. In general, the degree of polarization of most astrophysical sources is only a few percent. Therefore, it is essential to achieve the highest possible accuracy in polarimetric measurements. A simple approach to polarimetry involves rotating a polarizer and acquiring multiple exposures at different angles, from which the degree of linear polarization can be derived. However, in such systems, variations in atmospheric conditions between exposures can significantly degrade the measurement accuracy, making the results highly sensitive to atmospheric instability. To overcome this limitation, conventional astronomical polarimeters commonly employ polarization beam splitters (e.g., Wollaston prisms, WPs) in combination with a phase retarder such as a half-wave plate (HWP). These instruments enable the simultaneous sampling of orthogonal polarization components, effectively eliminating the influence of short-term atmospheric fluctuations. As a result, such systems can achieve polarimetric precisions of the order of $\sim$0.15\% \citep[e.g.,][]{2014SPIE.9147E..4OA}.


However, this configuration, which uses a WP, entails several technical challenges. First, the angular separation between the ordinary and extraordinary rays produced by the WP is relatively small (typically less than $20^\circ$). As a result, a focal plane mask (sometimes referred to as a polarization mask in the literature) is required to prevent overlap between the two components, which in turn limits the effective field of view (FOV) of the instrument. In addition, WPs are relatively expensive and can impose a financial burden on instrument development budgets. Moreover, since WPs are fabricated by cementing two birefringent crystals together, they may suffer from durability issues under thermal stress.


In light of these limitations, our research group has developed a low-cost polarimetric system with a moderately wide FOV that does not employ a WP. We thus designed the new polarimeter, the Seoul National University QUadruple Imaging Device for POLarimetry (SQUIDPOL). This instrument employs a non-polarizing beam splitter (NPBS), four wire-grid filters (WGFs), and four CMOS imagers. WGFs are more durable and cost-effective than WPs, offering a large (i.e., $90^\circ$) split angle, eliminating the need for a focal-plane mask, and enabling a wider FOV. This configuration enables the simultaneous measurement of the intensity of incident light in four polarization directions, allowing for the derivation of all four Stokes parameters in a single exposure. The recent reduction in the cost of high-performance commercial CMOS cameras has significantly contributed to the feasibility of our design. In conventional polarimetric systems, detectors were typically the most expensive component, and both the ordinary and extraordinary rays from the WP were projected onto a single detector to minimize the development cost. In contrast, our system leverages the affordability of modern CMOS sensors, enabling each polarization channel to be recorded with its own dedicated detector.

\begin{figure}[t]
\centering
\includegraphics[angle=0,width=80mm]{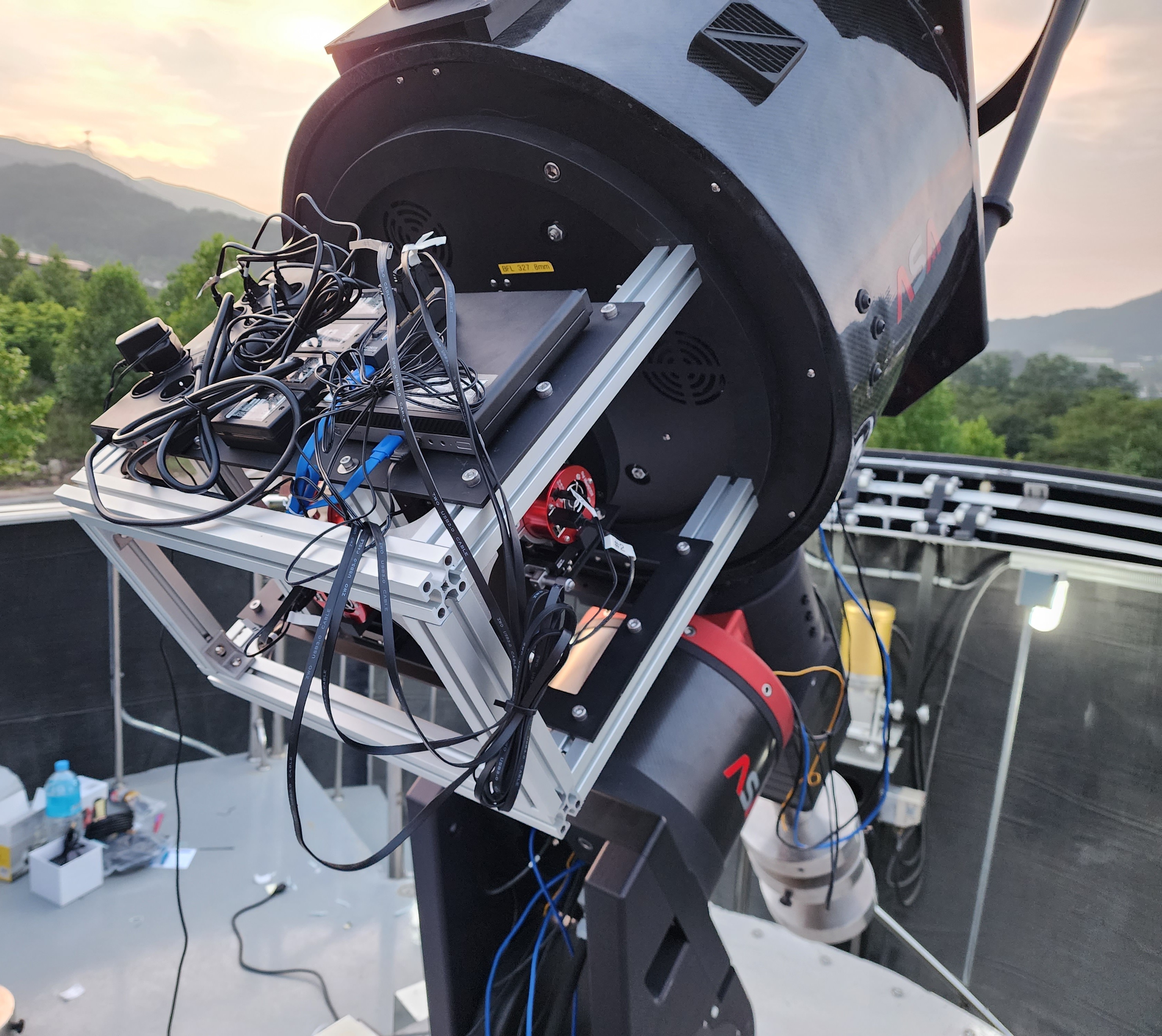}
\caption{SQUIDPOL mounted on the 60-cm telescope at the Pyeongchang Observatory of SNU. The instrument is enclosed by an aluminum frame that supports the optomechanical structure.\label{fig:1}}
\end{figure}

We developed SQUIDPOL at the Gwanak Campus of Seoul National University (SNU) until early 2024, and installed it on the 60-cm Ritchey-Chr\'etien telescope (the focal length of 4200 mm, f/7) at the Pyeonchang Observatory of SNU (Fig. \ref{fig:1}) in 2024 July. Since its installation, SQUIDPOL has been used for observations of Solar System objects, particularly comets, and the resulting scientific outcomes have been published \citep{2025ApJ...983L..19L,2026arXiv260108591C}.
In this paper, we present the optical and optomechanical design of SQUIDPOL Section \ref{sec:design}. We then report the performance results based on laboratory tests and on-site observations (Section \ref{sec:performace}). 

\begin{figure*}[t]
\centering
\includegraphics[angle=0,width=180mm]{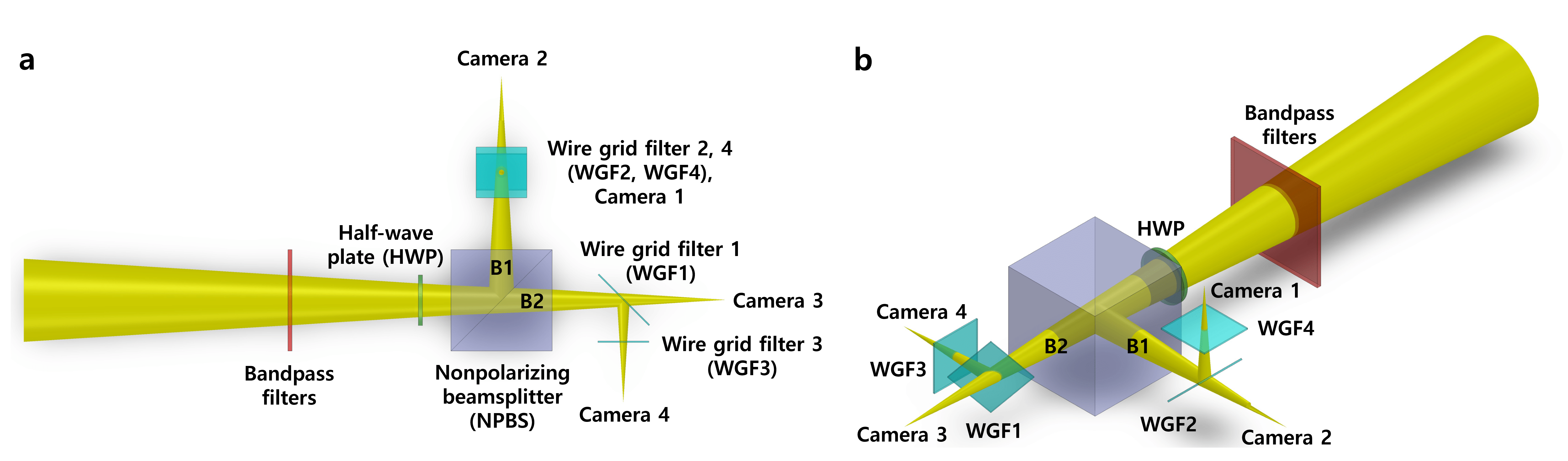}
\caption{Optical layout of SQUIDPOL. (a) Top view from the +X direction. (b) Isometric view. B1 is the first branch reflected from the NPBS, and B2 is the second branch, transmitted through the NPBS. \label{fig:2}}
\end{figure*}

\section{Instrument Design\label{sec:design}}

In this section, we describe the optical and optomechanical designs as shown below.

\subsection{Optical layout\label{sec:optics}}
Figure \ref{fig:2} illustrates the optical layout of SQUIDPOL. Bandpass filters are the first elements in the optical system. We installed an FLI Centerline CL1-10 filter wheel, which accommodates five 50 $\times$ 50 mm square filters in each of its two internal wheels. One wheel (wheel 0) contains Johnson-Cousins $B$-, $V$-, $R_\mathrm{C}$-, and $I_\mathrm{C}$-band filters, while the other wheel (wheel 1) holds a shutter, a WGF, and a $U$-band filter. This configuration allows for polarimetric observations and calibration using the WGF in the $B$-, $V$-, $R_\mathrm{C}$-, and $I_\mathrm{C}$-bands, photometry in the $U$-, $B$-, $V$-, $R_\mathrm{C}$-, and $I_\mathrm{C}$-bands, and the acquisition of bias and dark frames using the shutter. We use the 27105 Classic UBVRI filter set from Chroma Technology Corp. The filters have central wavelengths and bandwidth (FWHM) of 0.360 $\mu$m (0.066 $\mu$m), 0.440 $\mu$m (0.094 $\mu$m), 0.550 $\mu$m (0.088 $\mu$m), 0.640 $\mu$m (0.138 $\mu$m), and 0.790 $\mu$m (0.149 $\mu$m) for the $U$, $B$, $V$, $R_\mathrm{C}$, and $I_\mathrm{C}$ bands, respectively.


Following the bandpass filters, an Edmund Optics achromatic half-wave plate (HWP) (\#39-033) is positioned to rotate the polarization ellipse of the incoming light with respect to its fast axis. The HWP is mounted on a THORLABS ELL14 rotation stage, which uses a resonant piezo motor to rotate optical elements with a resolution of 44 $\mu$rad. By rotating the HWP to 0$^\circ$ or 45$^\circ$, we can rotate the polarization direction of the incident beam to 0$^\circ$ or 90$^\circ$, respectively, effectively switching the camera that detects different components of polarized light from the same WGF (i.e., Camera~1 and Camera~2). This setup enables flat-field correction for each camera.

Immediately following the HWP, the Edmund Optics non-polarizing beam splitter (NPBS) (\#49-004) is placed. It is a glass cube that splits incident light into two perpendicular optical paths without affecting its polarization state.

\begin{table}[h]
    \centering
    \setlength{\tabcolsep}{2.5pt}
    \caption{Transmittances and Reflectances of Wire Grid Filters (WGFs) based on our laboratory tests}
    \label{tab:wgfs}
    \small
    \begin{tabular}{lccc}
        \toprule
        Product name & T$^{a}$ & R$^{b}$ & T/R \\
        \midrule
        Edmund Optics \#46-636 & 1.538$\times10^8$ & 1.670$\times10^8$ & 0.921 \\
        Edmund Optics \#48-545& 2.459$\times10^7$ & 2.492$\times10^7$ &  0.987 \\
        Meadowlark Optics (order-made) & 1.968$\times10^7$ & 2.615$\times10^7$  & 0.753 \\
        \bottomrule
    \end{tabular}
    \tabnote{$^{a}$Aperture sum of the light transmitted through the WGFs in electron units, $^{b}$ aperture sum of the light reflected by the WGFs in electron units}
\end{table}

The first path (B1), which is reflected at 90$^\circ$ by the NPBS, encounters a wire-grid filter (WGF2), a glass plate with metal wires aligned in a specific direction. WGF2 reflects light polarized parallel to the wires and transmits light polarized perpendicular to the wires. The reflected and transmitted beams are directed to two separate cameras (Camera~1 and Camera~2) that detect orthogonal polarization components, $I_0$ and $I_{90}$, respectively. The second optical path, which is transmitted through the NPBS, encounters another wire-grid filter (WGF1) rotated by 45$^\circ$. Two cameras (Camera~3 and Camera~4) positioned after WGF1 detect orthogonal polarization components, $I_{45}$ and $I_{135}$. The values of 0, 90, 45, and 135 indicate the angles between the polarization direction measured by each camera and the instrument's x-axis.

We selected Edmund Optics 25-mm Square Broadband Polarizing Beam Splitter (\#48-545) for use as the WGF, as it exhibited the smallest difference in reflectance and transmittance for unpolarized light among the polarizers we tested (Table \ref{tab:wgfs}).

Furthermore, we installed two additional WGFs (WGF3 and WGF4). These WGFs are placed after WGF1 and WGF2, respectively. They are positioned in front of Camera~1 and Camera~4, which receive the light reflected from WGF1 and WGF2. These additional WGFs filter out light reflected from the glass side (back side) of the primary WGFs, which is polarized perpendicular to the light reflected from the wire side (front side). For WGF3 and WGF4, we used 25-mm square custom-made WGFs produced by Meadowlark Optics, allowing us to orient the wires diagonally. This enabled a simpler and more compact design for the filter holder.

For the detectors, we employed four ZWO ASI 294MM cooled CMOS cameras, each with a sensor size of 19.4 $\times$ 13.0 mm. 
The cameras are synchronized to ensure simultaneous exposure. They are operated in 4$\times$4 binning mode, resulting in an effective pixel size of 9.2 $\mu$m, which corresponds to a pixel resolution of $0.45''$ when mounted on the 60-cm Ritchey-Chr\'etien telescope (i.e., the telescope with a focal length of 4200 mm). The notation of the cameras (Camera~1 through Camera~4) is also indicated in Fig. \ref{fig:2}.

\begin{figure*}[t]
\centering
\includegraphics[angle=0,width=180mm]{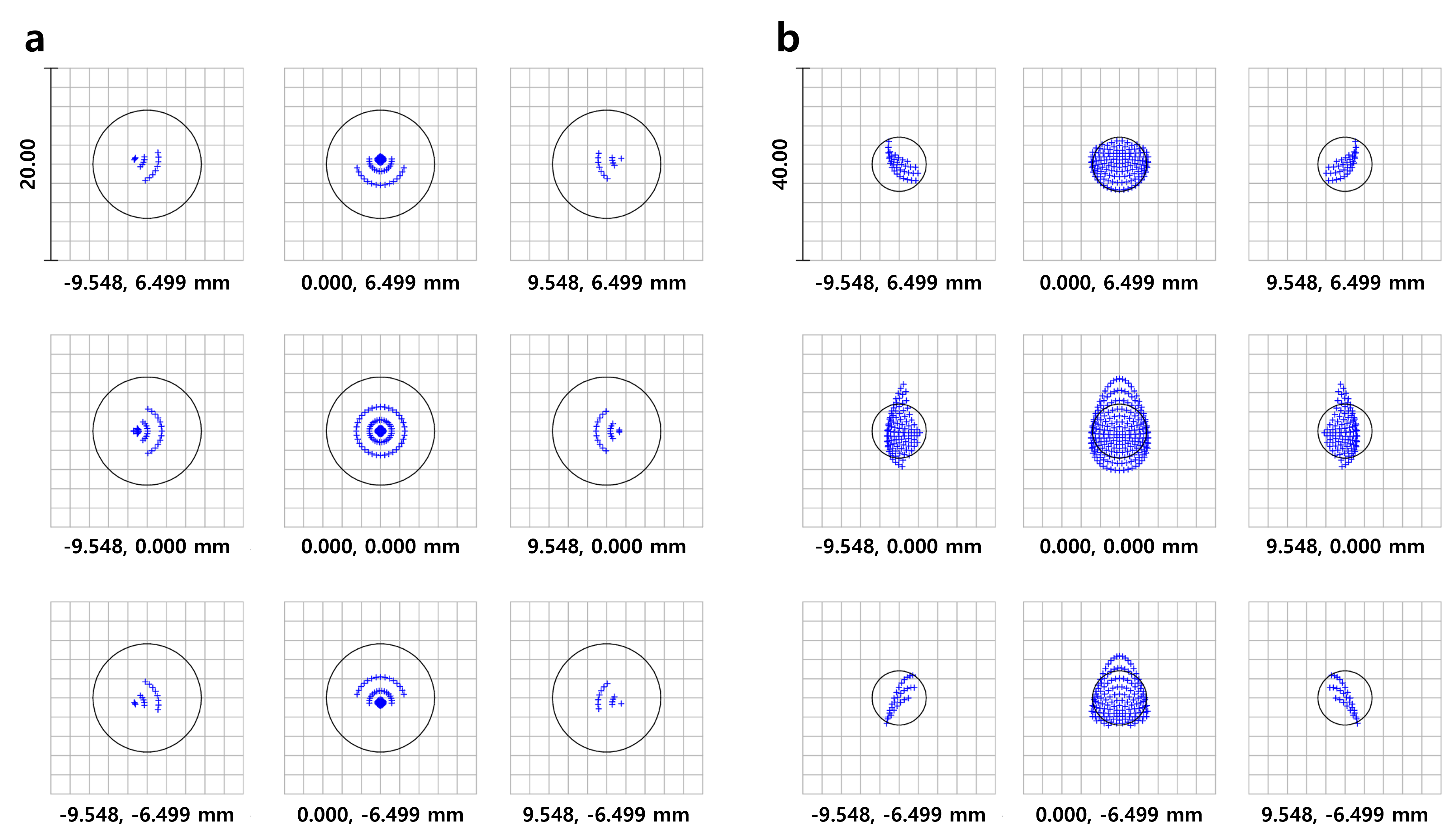}
\caption{ ZEMAX spot diagrams for (a) Camera~1 and (b) Camera~3. The numbers below each diagram indicate the X and Y positions (in mm) on the imaging plane. The overlaid circles represent the Airy disk size for the 60-cm telescope, with a radius of 4.7 $\mu$m at the V-band (550 nm). The spot diagrams for Camera~2 and Camera~4 are similar to those for Camera~3 and Camera~1, respectively, due to their identical optical layouts. \label{fig:3}}
\end{figure*}

We evaluated the optical performance of the system using ZEMAX. First, we optimized the positions of each sensor to minimize the radii of $80$\% encircled energy and analyzed the resulting spot diagrams (Fig. \ref{fig:3}). The spot diagram for Camera~1 shows diffraction-limited performance, while that for Camera~3 exhibits a larger spot size due to astigmatism introduced as the converging beam passes through WGF1, which is tilted at $45^\circ$. Nevertheless, the radii of the $80$\% encircled energy for Cameras~2 and 3 remain smaller than one pixel ($9.2 ~\mu$m, $0.45''$) across all operation wavelengths. This is acceptable for ground-based observations at the observatory, where the average seeing is expected to exceed $1''$.

\begin{table}[h]
    \centering
    \setlength{\tabcolsep}{2.5pt}
    \caption{Focal shifts due to chromatic aberration relative to the $V$-band (mm)}
    \label{tab:chromatic}
    \small
    \begin{tabular}{cccc}
        \toprule
        Camera No. & $B$ (445 nm) & $R_{\mathrm{c}}$ (658 nm) & $I_{\mathrm{c}}$ (806 nm) \\
        \midrule
        1 & 0.173 & -0.100 & -0.186 \\
        3 & 0.170 & -0.100 & -0.186 \\
        \bottomrule
    \end{tabular}
    \tabnote{The "+" sign indicates an increase in the backfocus distance.}
\end{table}

Additionally, the use of a converging beam introduces chromatic aberration, resulting in focal position shifts. We summarize the shift in Table \ref{tab:chromatic}. These shifts can be managed by adjusting the focus position of the 60-cm telescope. Moreover, we noticed that this system exhibits noticeable vignetting. It is mainly caused by the limited apertures of HWP, WGF1, and WGF2. This effect can be corrected through flat-fielding and by using multiple HWP angles. However, we plan to eliminate this vignetting effect
by enlarging these optical element sizes.

\subsection{Optomechanical  design\label{subsec:tolerancing}}

We performed a tolerancing analysis using ZEMAX to estimate the degradation of optical performance due to alignment errors. In this paper, we adopt the coordinate system defined by \citet{Kim2010}, where $\alpha$-, $\beta$-, and $\gamma$-tilts correspond to rotations around the x-, y-, and z-axes, respectively. As the criterion for tolerancing, we analyzed the boresight errors and ensured that the FOV centers of the four detectors deviate from the nominal position by less than $1'$. Here, boresight error is defined as the displacement of the FOV center on the image plane from its nominal position. We selected Camera~1 for the analysis because it receives light reflected by both the NPBS and WGF2, whose tilts are the dominant contributors to the boresight error. In this analysis, we considered only the NPBS, WGF2, and WGF4 because they will be mounted in newly designed holders and are expected to introduce the most significant boresight errors in Camera~1. Table \ref{tab:tolerance} summarizes the tolerances considered in our analysis.

\begin{table}[h]
    \centering
    \setlength{\tabcolsep}{2.5pt}
    \caption{Tolerances considered for SQUIDPOL. 
    }
    \label{tab:tolerance} 
    \small
\begin{tabular}{llr}
\toprule
Optical Element       & Parameter                      & Value                   \\
\midrule
\multirow{4}{*}{NPBS} & X and Y decenter (each)                  & $\pm$ 0.1 mm                  \\
                      & Despace from HWP               & $\pm$ 0.1 mm                  \\
                      & $\alpha$ and $\beta$ tilt (each)              & $\pm$ 0.25$^\circ$                \\
                      & $\gamma$ tilt                     & $\pm$ 0.5$^\circ$                 \\
\midrule
\multirow{4}{*}{WGF2} & X and Y decenter (each)                  & $\pm$ 0.1 mm                  \\
                      & Despace from NPBS              & $\pm$ 0.1 mm                  \\
                      & $\alpha$ and $\beta$ tilt (each)              & $\pm$ 0.3$^\circ$                 \\
                      & $\gamma$ tilt                     & $\pm$ 0.5$^\circ$                 \\
\midrule
\multirow{4}{*}{WGF4} & X and Y decenter (each)                 & $\pm$ 0.1 mm                  \\
                      & Despace from WGF2              & $\pm$ 0.1 mm                  \\
                      & $\alpha$ and $\beta$ tilt (each)               & $\pm$ 0.5$^\circ$                 \\
                      & $\gamma$ tilt                     & $\pm$ 0.5$^\circ$                 \\
\bottomrule
\end{tabular}
\end{table} 

\begin{figure*}[t]
\centering
\includegraphics[angle=0,width=180mm]{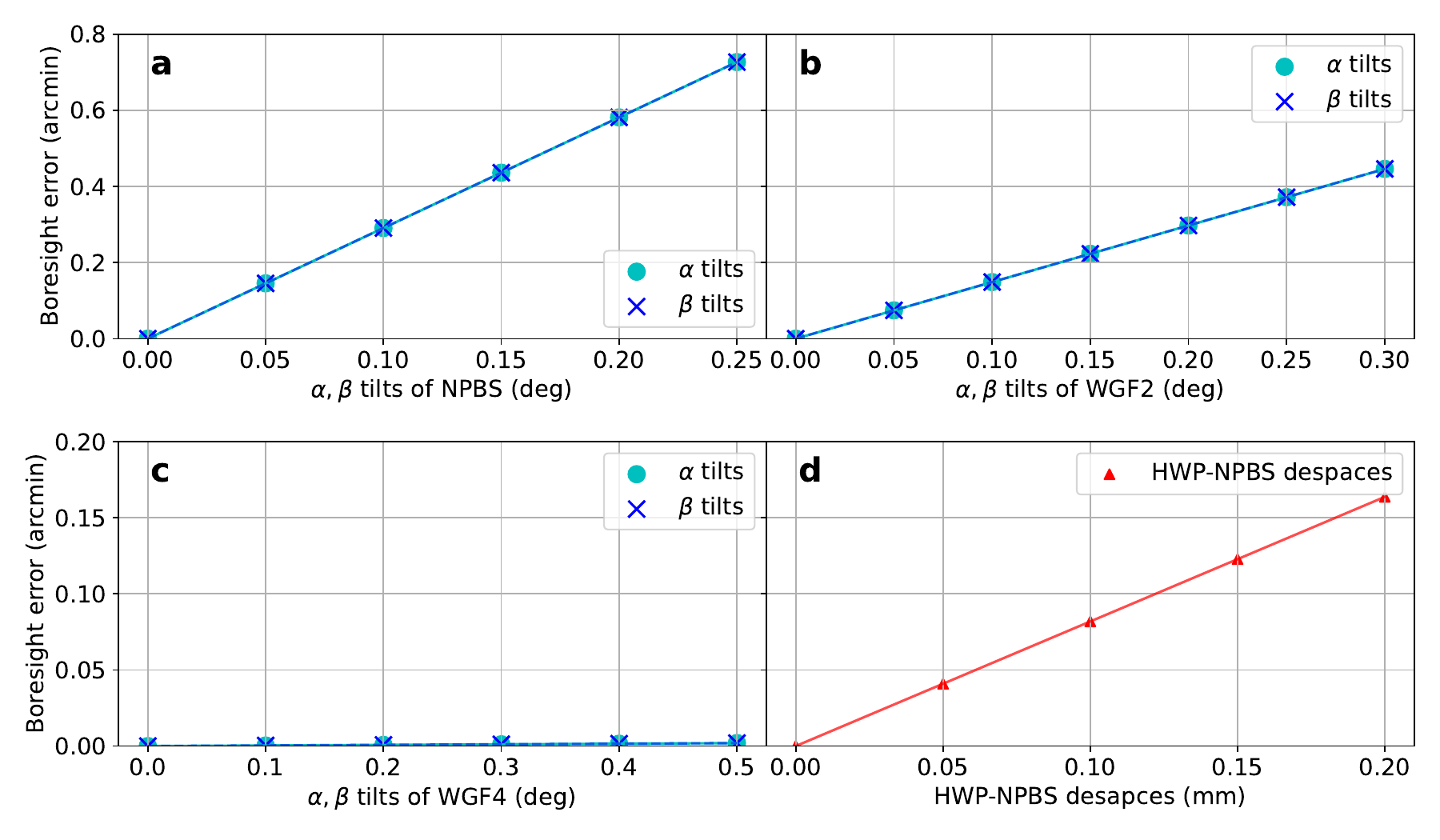}
\caption{Sensitivity analysis results. (a)--(c) Boresight errors caused by $\alpha$-tilts (cyan circles) and $\beta$-tilts (blue crosses) of the (a) non-polarizing beam splitter (NPBS), (b) second wire-grid filter (WGF2), and (c) fourth wire-grid filter (WGF4). (d) Boresight errors resulting from despace between the HWP and NPBS. Other tolerances are not shown, as their effects on the boresight error of Camera~1 are smaller than $1"$. \label{fig:4}}
\end{figure*}

\begin{figure}[]
\centering
\includegraphics[angle=0,width=80mm]{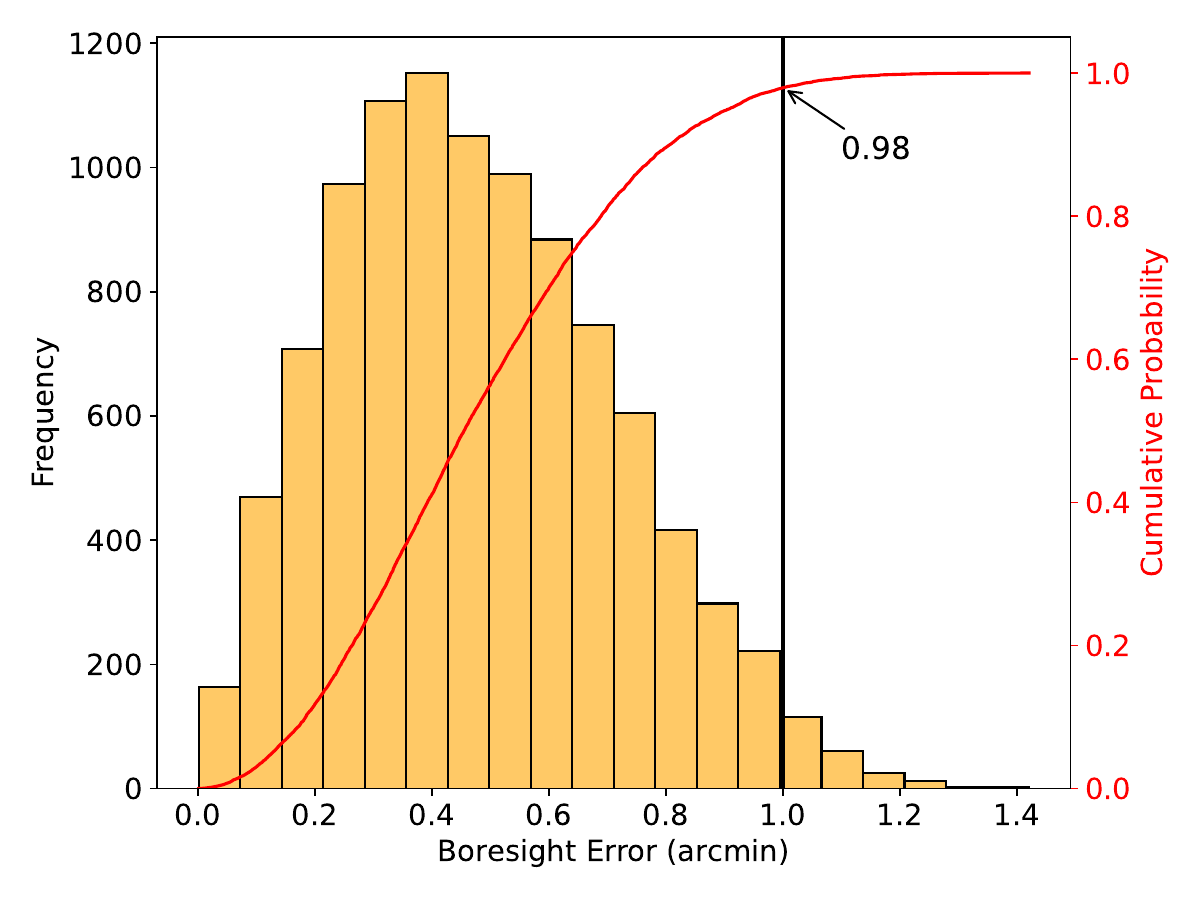}
\caption{Probability distribution of boresight errors from 10,000  Monte Carlo simulations. The vertical black line at $1'$ indicates the tolerance limit. The cumulative probability (red solid line) shows that 98\% of the simulations meet the criterion. \label{fig:5}}
\end{figure}

Figure \ref{fig:4} is the result of the sensitivity analysis. We confirmed that $\alpha$- and $\beta$-tilts of both the NPBS and WGF2 are the most sensitive parameters for Camera~1. On the other hand, other parameters do not affect the boresight error by more than $1'$. Next, we performed a Monte Carlo analysis for Camera~1. Each simulation evaluates the boresight error using the parameters listed in Table \ref{tab:tolerance}, assuming they follow Gaussian distributions with the specified tolerance values as their standard deviations. We determined the tolerance limits that satisfy our criterion, boresight errors smaller than $1'$, in 98\% of 10,000 simulations (Fig. \ref{fig:5}).

Following these tolerances, we designed dedicated holders for the NPBS and WGFs. In particular, the stringent requirements on $\alpha$- and $\beta$-tilts required special consideration due to their significant influence on boresight errors. Therefore, we carefully engineered the holder designs to minimize these tilts.

The NPBS is secured by the filter box using a 3-2-1 positioning scheme to ensure accurate alignment. Specifically, three and two protrusions (each 2 $\times$ 4 mm$^2$ and 2 mm high) are positioned on the filter holder along the +Z and +X directions, respectively. A Teflon-padded bolt is tightened from the -Y direction to complete the constraint system. We specified the tolerance of the protrusion distances to be 0.05 mm, half the value of other dimensional tolerances (0.1 mm), to further minimize potential tilts.

On the opposite faces, we installed three, two, and four M4 ball plungers, respectively, each with a minimum elastic force of 5 N. This configuration ensures that even with only two plungers engaged, the holding force (10 N) exceeds three times the gravitational force on the NPBS (314 g), providing sufficient stability.

WGF1 and WGF2 are mounted using filter holders that attach to the main filter box. These holders also employ a 3-2-1 positioning scheme to accurately align the WGFs relative to the holder. Additionally, locating pins are used to securely position each filter holder with respect to the filter box.

We also conducted a stability analysis of the system using the static analysis function of SOLIDWORKS to assess possible deformations caused by gravity that could shift the location of the imaging plane. We assumed the system was fixed to the telescope, applied gravity in the +X, -X, +Y, -Y, +Z, and -Z directions, and calculated displacements from the original positions along the X, Y, and Z axes. To quantify the resulting image shifts, we then used the probe function to extract decenter and defocus values of the imaging planes. We calculated average values and standard deviations from 9 nodes within each imaging plane. 

We found that the largest average decenters of Camera~1, Camera~2, Camera~3, and Camera~4 from the static analysis are $2.643\pm0.081$, $3.958\pm0.178$, $6.915\pm0.025$, and $5.803\pm0.192~\mu\mathrm{m}$, respectively. These values ensure the stability of the system during a full night of observation, as they are smaller than the effective pixel size from 4 $\times$ 4 binning (9.2 $\mu$m). On the other hand, the absolute values of the largest average defocuses from the same analysis are estimated to be $4.002\pm0.107$, $2.613\pm0.131$, $0.681\pm0.056$, and $3.890\pm0.106~\mu\mathrm{m}$ for Cameras~1--4, respectively. Considering the F-ratio of the 0.6-m telescope (F/7), these defocuses are expected to broaden the FWHM to approximately $0.03''$. Because this value is significantly smaller than the typical seeing size ($\sim1.5''--2.0''$), we suggest that such small defocuses are negligible for the actual observations.

\section{Performance evaluation results\label{sec:performace}}

In this section, we present the performance evaluation results based on our test observations conducted with the 0.6-m telescope.
These evaluations aim to assess whether SQUIDPOL meets the design requirements for high-precision polarimetric observations.


\subsection{Test Observations}\label{subsec:observations}
Test observations were carried out over 8 nights from 2024 September to 2026 January using the 0.6-m telescope at Pyeongchang Observatory of SNU 
The observations were performed with $B$-, $V$-, $R_\mathrm{C}$-, and $I_\mathrm{C}$-filters. The observed targets and their corresponding purposes are summarized in Table \ref{tab:test_obs}.

\begin{table}[h]
\centering
\small
\setlength{\tabcolsep}{2pt}
\caption{Summary of Test Observations}
\label{tab:test_obs}
\begin{tabular}{lll}

\toprule
Category & Target & Date (YYMMDD) \\
\midrule
Unpolarized Stars$^{a}$ & Caph & 251230, 260119 \\
                        & $\gamma$~Boo & 260116 \\
Strongly Polarized Stars$^{b}$ & HD25443 & 251225,251229, \\
&&260119 \\
                               & HD7927 & 251229, 260116 \\
100\% Polarized stars$^{c}$ & Bright Star (+WGF) & 251229, 260119 \\
  &  & 260121 \\

Open Cluster$^{d}$ & M34 & 240926 \\
 & M29 & 241017 \\
 & NGC 1502 & 251103\\
 & M 36 &251103\\
\bottomrule
\end{tabular}
\tabnote{

$^{a}$ Instrumental polarization calibration. 
$^{b}$ Polarization angle offset and consistency check with catalog values. 
$^{c}$ Instrumental polarization efficiency, assuming 100\% polarized incident light when the WGF is placed in filter wheel~1. 
$^{d}$ Camera FOV offsets (X, Y, and rotation) and system stability. Dates are given in UT.
}
\end{table}

\subsection{Geometric characteristics}\label{subsec:geometry}
Because SQUIDPOL comprises four cameras, it is important to align their FOVs as closely as possible and to investigate their geometric relationships. Although further fine-tuning remains possible, this section summarizes our current efforts to characterize the geometric performance of the SQUIDPOL system at its present stage. 

We investigated the geometric characteristics by using the solve-field command of Astrometry.net to determine World Coordinate System (WCS) solutions for each image \citep{2010AJ....139.1782L}. A total of 48 images (12 per camera) of the open cluster M11, taken on 2024 October 24 UT, were used to determine the width and height of the FOV, pixel scale, and offsets in image centers and rotations between cameras. We calculated offsets with respect to Camera~3, as its optical path does not include any reflective elements that could introduce boresight errors, as discussed in Section \ref{subsec:tolerancing}.

\begin{figure}
\centering
\includegraphics[angle=0,width=80mm]{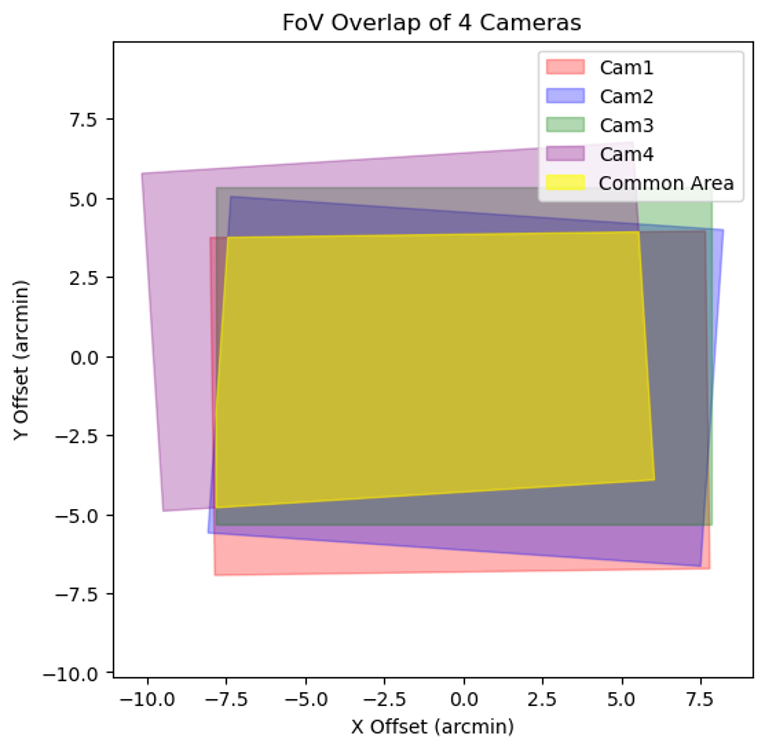}
\caption{Field of view of the four cameras with respect to Camera~3. \label{fig:n}}
\end{figure}

\begin{table*}[t!]
\caption{Astrometric Performance\label{tab:astrometry}}
\centering
\begin{tabular}{ccccccc}
\toprule
Camera No. & Width & Height & Pixel scale & X offset & Y offset& Rotational offset \\
 &(arcmin) &(arcmin) &(arcsec/pix) &(arcmin) &(arcmin) &(deg) \\
\midrule
1 & 15.659(7) & 10.676(7) & 0.4537(1) & -0.105(2) & -1.494(3) & 0.766(18) \\
2 & 15.616(8) & 10.662(7) & 0.4528(2) & 0.074(2) & -0.803(3) & -3.865(41) \\
3 & 15.642(7) & 10.670(4) & 0.4533(2) & - & - & - \\
4 & 15.568(16) & 10.699(23) & 0.4529(7) & -2.065(3) & 0.926(3) & 3.625(29) \\ 
\bottomrule
\end{tabular}
\tabnote{
Numbers in parentheses indicate uncertainties in the last digit(s).\\
}
\end{table*}

Table \ref{tab:astrometry} summarizes the astrometric performance of SQUIDPOL. The filter box was designed to limit boresight errors between cameras to less than $1'$.
Nevertheless, Camera~1 and Camera~4 exhibit Y and X offsets exceeding this threshold, respectively.
These deviations can be interpreted as a consequence of the fact that both cameras receive reflected light from the WGFs, which enhances the effect of small misalignment tilts in the holders of WGF1 and WGF2.
Figure \ref{fig:n} shows the areas covered by each camera with respect to the Camera~3 position. Despite the offsets, the common area observed by all four cameras is 110.6 arcmin$^2$.
Despite minor misalignments, the current SQUIDPOL retains sufficient overlap among the four cameras to conduct polarimetric measurements across moderately large common FOVs.

\subsection{Optical characteristics}
\label{subsec:optical}

\begin{figure}[htbp]
\centering
\includegraphics[width=0.48\textwidth]{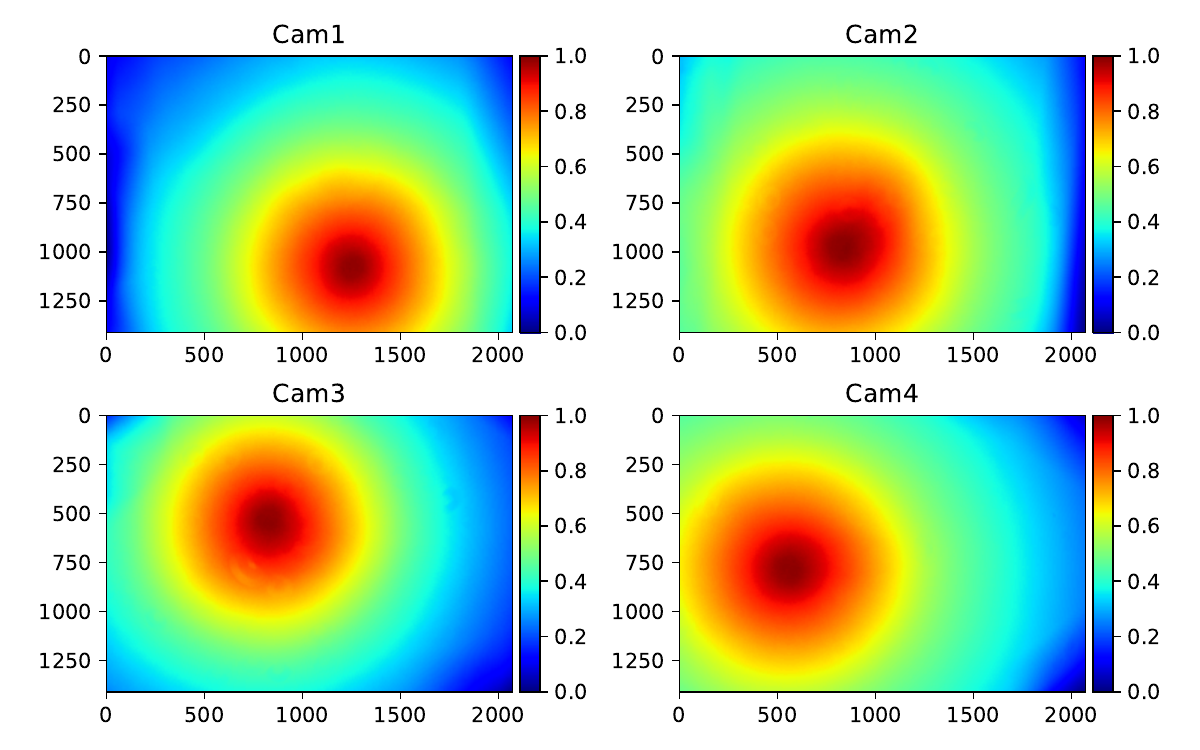}%
\hfill
\includegraphics[width=0.48\textwidth]{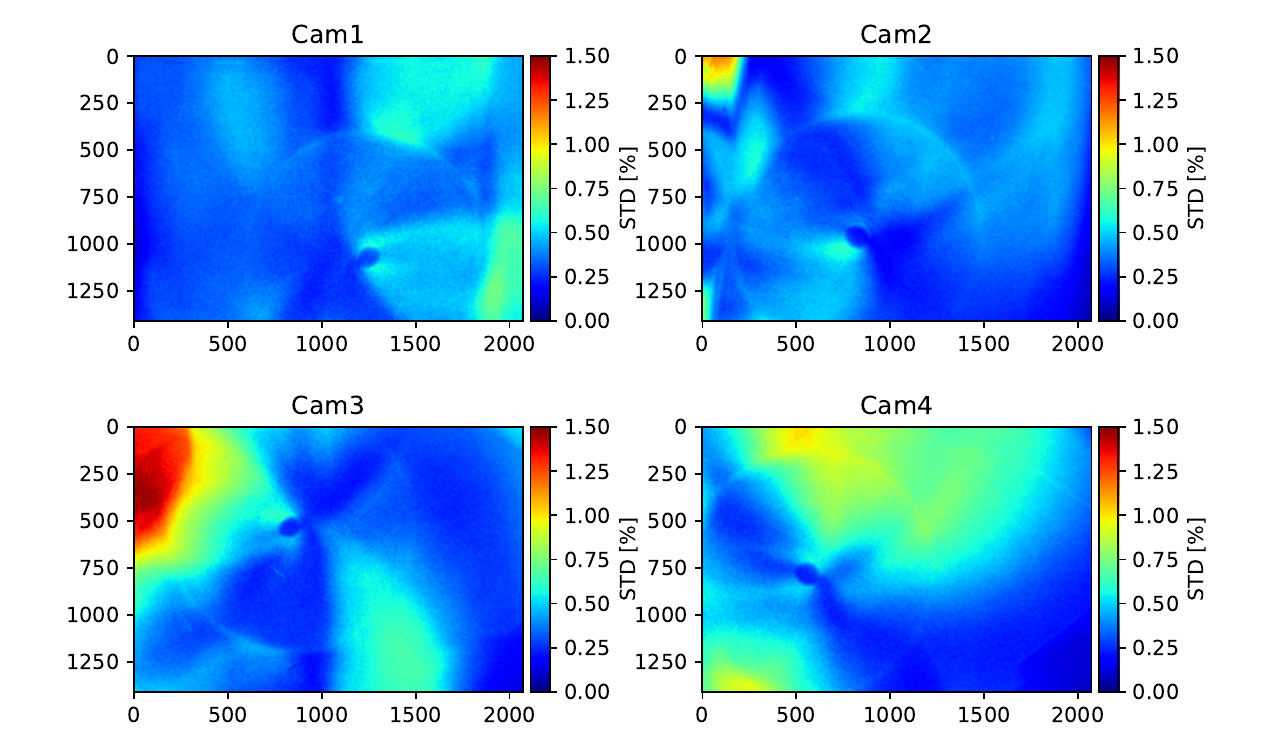}
\caption{(Top four panels) Twilight flat-field images normalized to the peak intensity. 
(Bottom four panels) Pixel-wise standard deviation maps derived from four flat-field images taken at HWP angles of 0$^\circ$, 45$^\circ$, 90$^\circ$, and 135$^\circ$.
In each row, the panels correspond to the four cameras (Camera~1--Camera~4).}
\label{fig:flat_std}
\end{figure}

\begin{figure}[htbp]
\centering
\includegraphics[width=0.88\linewidth]{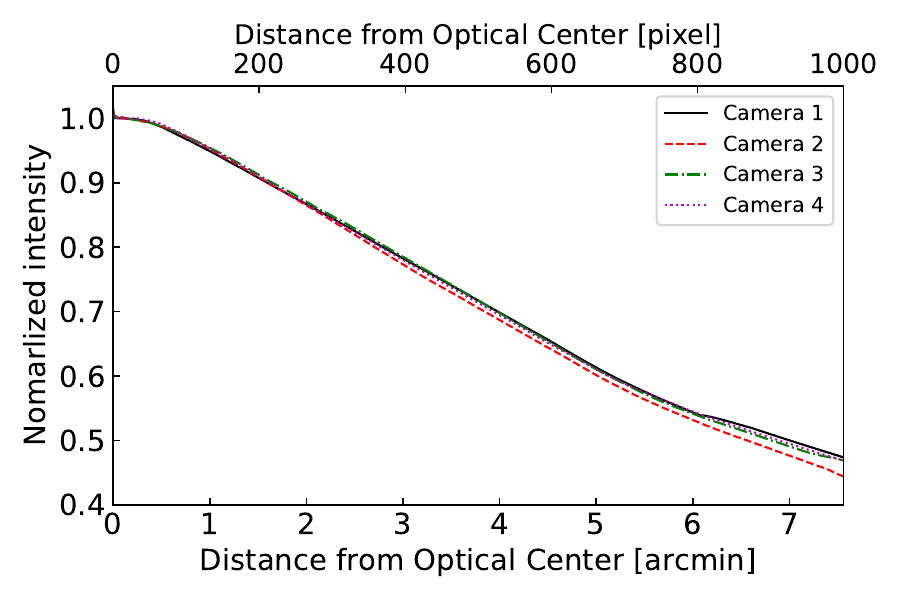}
\caption{Normalized flat-field intensity profiles as a function of distance from the optical center.}
\label{fig:flat_profile}
\end{figure}

As we mentioned in Section \ref{sec:optics}, SQUIDPOL utilizes off-the-shelf optical components, including a HWP with a limited aperture size. According to our estimates using ZEMAX ray-tracing simulations, the current design of SQUIDPOL is expected to cause significant vignetting, with an estimated 50\% decrease in intensity at $\sim 7.4'$ away from the optical center.
Figures \ref{fig:flat_std} show twilight flat-field images taken near the zenith direction (top) and the corresponding pixel-wise standard deviation maps (bottom). In Figure \ref{fig:flat_std} (top), we normalized the intensity values at the brightest regions to unity. In Figure \ref{fig:flat_std} (bottom), we calculated these standard deviation maps using four flat images taken at different HWP angles of 0$^\circ$, 45$^\circ$, 90$^\circ$, and 135$^\circ$, in order to evaluate the variation at each pixel position. 
Clear vignetting is observed in Figure \ref{fig:flat_std} (top).
To make a better comparison among the four camera data, we made Figure \ref{fig:flat_profile}, where the normalized flat-field intensities are given as a function of distance from the optical center (i.e., the brightest region). From Figure \ref{fig:flat_profile}, approximately 50\% vignetting at the periphery, around $5'$ from the center, is observed. 
Furthermore, no significant variation is observed among the four cameras (Camera~1 to Camera~4), indicating that the vignetting originates from an optical element located upstream of the non-polarizing beam splitter. This result is consistent with the ZEMAX simulation, which predicts optical obscuration caused by the limited aperture of the HWP. We therefore conclude that the small aperture of the HWP is the limiting factor responsible for this severe vignetting. 
At present, the vignetting effect is corrected during data reduction by acquiring flat-field images (either sky flats or twilight flats) and applying appropriate corrections.

We also note that a small flat-field variation with an amplitude of $\lesssim$ 0.5\% is observed as the HWP rotates (bottom four panels in Figure~\ref{fig:flat_std}). The variations exhibit ring-like structures and surrounding features, although their physical origin is currently unclear. In contrast, the flat-field variation near the optical center is much smaller, at a level of $\ll$ 0.25\%, indicating that observations conducted close to the optical axis are less affected by this effect.

As we explain later (in Sect \ref{subsec:stability}), the influence of flat-field variations would be canceled out when we derive the polarization degrees using SQUIDPOL data. Therefore, we consider that the current optical configuration is sufficient to achieve reliable polarimetric accuracy. Nonetheless, if future funding permits, we plan to introduce a larger-aperture HWP to improve the signal-to-noise ratio across the full field of view, including the outer regions.

The point spread function (PSF) of SQUIDPOL was determined using observations of the open clusters M29, NGC 1502, and M36 conducted on 2024 October 17, and 2025 November 03 UT. A total of 236 images (59 per camera) were obtained. The observations were carried out in the $R_\mathrm{C}$ band with the HWP set to 0$^\circ$. The PSF for each camera was measured by fitting a 1D Gaussian function to the point sources in the images. The numbers of point sources used for the fitting were 1815, 2025, 2301, and 1521 for cameras~1, 2, 3, and 4, respectively. The main reason for the different numbers of point sources is the difference in FOV among the cameras. We then derived the average FWHM of the fitted Gaussian functions for each camera, applying 3-sigma clipping to exclude outliers. The resulting PSF FWHM mean values and their 1-sigma standard deviations are $3.25''\pm0.71''$, $3.39''\pm0.86''$, $3.32''\pm0.85''$, and $3.33''\pm0.84''$ (or $7.15\pm1.56$, $7.48\pm1.91$, $7.33\pm1.88$, and $7.35\pm1.85$ pixels) for Cameras~1, 2, 3, and 4, respectively. 
No significant variation in the PSF was found across the FOV of any individual camera.

\begin{figure}
\centering
\includegraphics[width=0.9\linewidth]{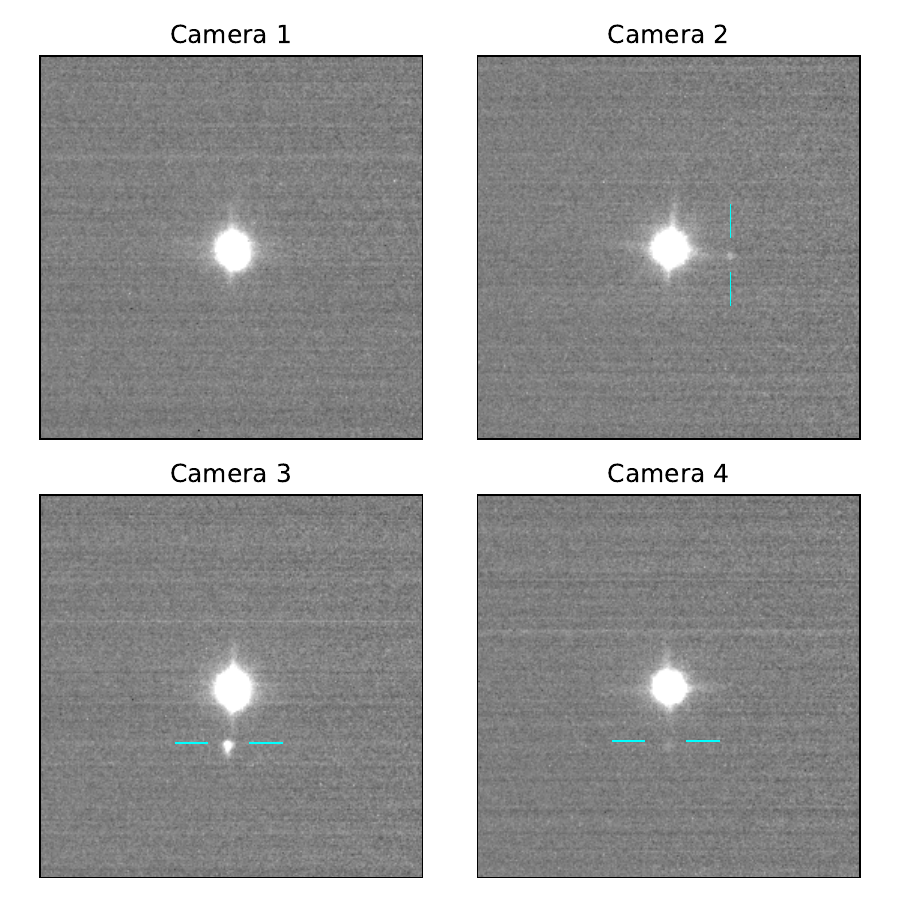}
\caption{Ghost image is seen near a bright target at the center of the image. The ghost images are located within the cyan-colored lines. The images are cropped to 2.7' $\times$ 2.7'. The target is the unpolarized standard star, $\gamma$~Boo}
\label{fig:Ghost}
\end{figure}

Figure \ref{fig:Ghost} shows ghost patterns visible near bright targets.  These ghost patterns are $\sim26.2''$  away from the real images, with their intensity varying by cameras. The aperture sum of the ghost images is 0.07\%, 0.3\%, and 0.05\% of that of the real images for Camera~2, 3, and 4, respectively. In Camera~1, the ghost image is undetectable as its intensity is nearly zero. Cameras~1 and 4 show lower ghost fractions because WGF3 and WGF4 effectively filter polarized light reflected from the back surfaces of WGF1 and WGF2, which are the primary sources of the ghost patterns in these cameras. We expect that the impact of these ghost patterns on the polarization results is negligible. 
For point sources, the separation between the ghost and the real image is several times larger than the typical seeing size at the observation site (typically $2''-3''$), placing the ghost pattern outside the sky annulus used for aperture photometry. We therefore confirm that the ghost pattern does not contribute to the polarization accuracy for point sources. On the other hand, for extended sources, ghost patterns could potentially cause a deviation in the measured polarization from the true value. However, we confirm that such an effect does not exceed 0.05\%$_\mathrm{p}$. This upper limit was derived by simulations of extended sources, incorporating the measured ghost offsets and flux ratios for all four cameras. By generating synthetic surface-brightness maps with and without ghost effects applied, we derived polarization maps. The comparisons show that the induced deviation remains below 0.05\%$_\mathrm{p}$, regardless of the input polarization degree.

\begin{table}[t!]
\caption{Gain and readout noise measured for each camera}
\label{tab:gainread}
\centering
\begin{tabular}{ccc}
\toprule
Camera No. & Gain & Readout Noise \\
 &(e$^-$/ADU) & (e$^-$)  \\
\midrule
1 & $3.59 \pm 0.23$ & $13.73 \pm 0.93$  \\
2 & $3.58 \pm 0.24$ & $13.71 \pm 0.93$  \\
3 & $3.41 \pm 0.36$ & $13.13 \pm 1.40$   \\
4 & $3.51 \pm 0.29$ & $13.44 \pm 1.12$  \\ 
\bottomrule
\end{tabular}
\end{table}

\subsection{Basic camera properties}
\label{subsec:camera}

Before evaluating the system stability and polarimetric performance of SQUIDPOL, we characterized the fundamental detector properties relevant to the following analyses.
We measured the gain, readout noise, and their temperature dependence for each camera.
Following Janesick’s method, we derived the conversion gain and readout noise from 79 flat-field image pairs and 15 bias image pairs \citep{1985SPIE..570....7J}.
To mitigate illumination gradients, all measurements used a central $99 \times 99$-pixel region.
The results are summarized in Table \ref{tab:gainread}.

\begin{figure}
\centering
\includegraphics[width=0.8\linewidth]{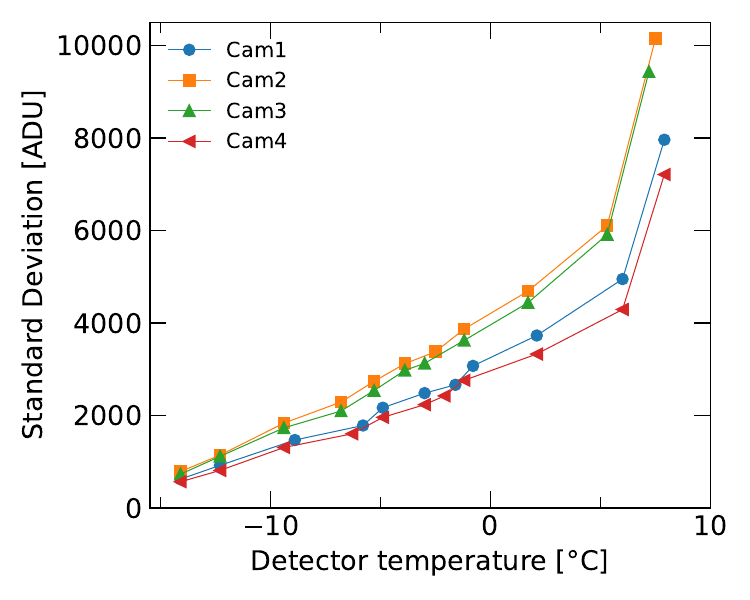}
\caption{Temperature dependence of the standard deviation of dark frames for the ZWO ASI 294MM CMOS detectors.}
\label{fig:DET-TEM}
\end{figure}

The temperature dependence of the detector was investigated.
Figure \ref{fig:DET-TEM} shows the correlation between the standard deviation (STD) of dark frames and the detector temperature for the ZWO ASI 294MM CMOS camera.
The STD decreases linearly as the sensor temperature drops from 5$^\circ$C to –15$^\circ$C.
However, when the temperature exceeds $\sim$5$^\circ$C, the STD increases significantly.
This indicates that operating the detector at temperatures above $\sim$5$^\circ$C is unfavorable for observations.

\subsection{System stability}
\label{subsec:stability}

\begin{figure}
\centering
\includegraphics[width=\linewidth]{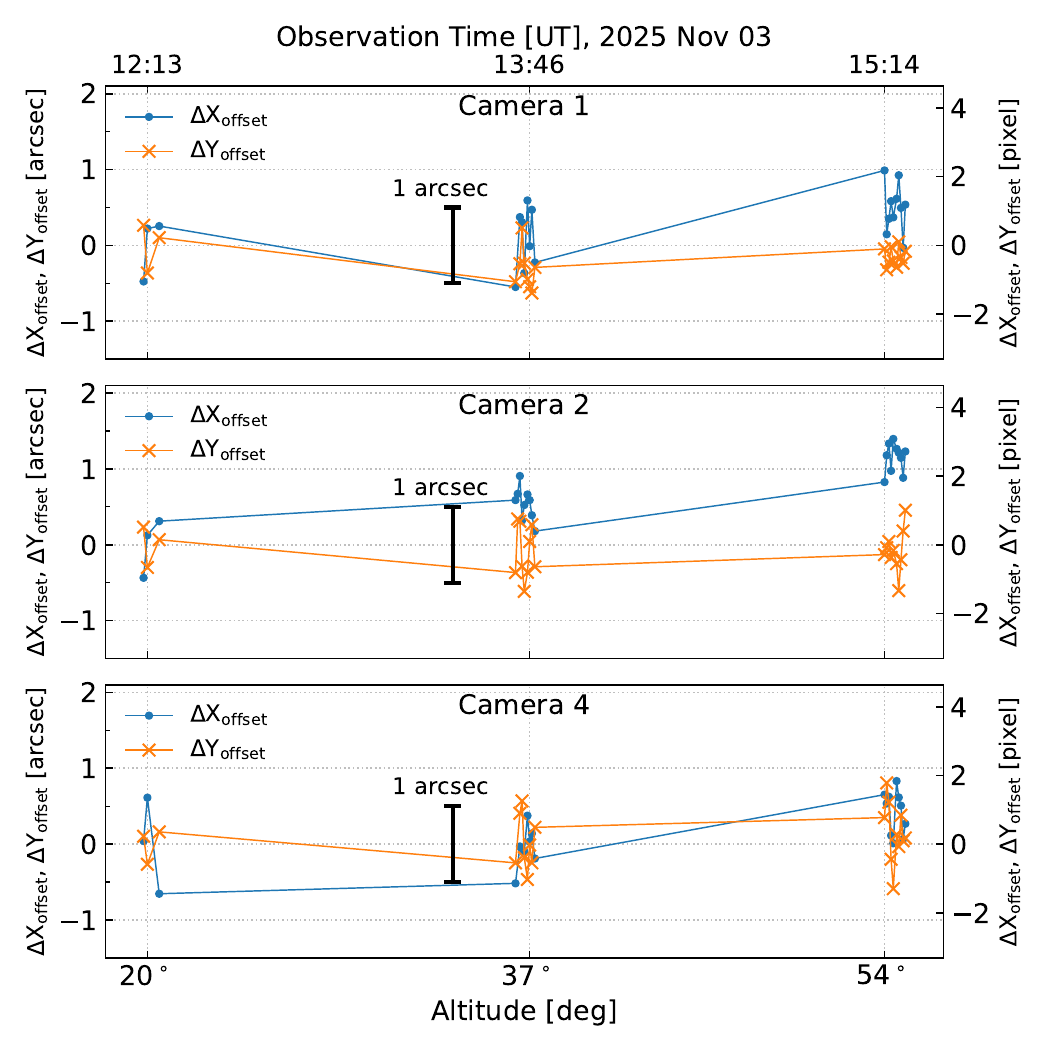}
\caption{Stability of each camera during the observation of the open cluster M36 on 2025 November 03 UT.
The telescope pointing direction varied over the observation, changing the gravity direction acting on the cameras. The changes in the X and Y offsets are shown relative to the mean offsets measured from the first three images of each camera. The left y-axis shows offsets in arcsec, and the right y-axis shows the corresponding values in pixels. A scalebar corresponding to 1 arcsecond is shown for reference. }
\label{fig:stability}
\end{figure}

To evaluate the stability of SQUIDPOL, we observed the open cluster M36 on 2025 November 3 UT at different telescope pointing directions, corresponding to altitudes of $20^\circ$, $37^\circ$, and $54^\circ$, which changed the direction of gravity acting on each camera. A total of 96 images (24 images per camera) were obtained in the $R_\mathrm{C}$ band with the HWP fixed at $0^\circ$. The celestial coordinates (right ascension and declination) of the image center for each camera were measured using the \texttt{solve-field} command of Astrometry.net \citep{2010AJ....139.1782L}. We then calculated the relative offsets of the image centers between the cameras, taking Camera~3 as the reference, and examined how these offsets depend on the telescope pointing direction. Figure~\ref{fig:stability} shows the temporal variations of the offsets for Cameras~1, 2, and~4 relative to Camera~3; Camera~3 itself is not shown since it serves as the reference. To emphasize the variations, the mean offsets in right ascension and declination measured from the first three images were subtracted from all data points.

Figure \ref{fig:stability} shows that the X and Y offsets of all cameras vary within approximately $1''$ when the telescope is pointed in the same direction
This variation corresponds to about two pixels in SQUIDPOL and is still significantly smaller than the typical seeing size at Pyeongchang ($2''-3''$). Therefore, the observed variations are more likely caused by atmospheric seeing and/or uncertainties in the \texttt{solve-field} solutions provided by Astrometry.net, rather than by physical movement of the cameras. Based on the static analysis performed in SolidWorks (Sect.~\ref{subsec:tolerancing}) and the observational measurements, we expect that camera movement due to changes in the gravity direction is negligible or remains smaller than the seeing-limited resolution.

Such stability implies that the flat-field pattern does not change its position between exposures.
If the flat-field pattern remains fixed, flat-field effects are expected to be largely reduced when deriving the polarization degree using SQUIDPOL data, as mentioned in Sect.~\ref{subsec:optical}. In SQUIDPOL, the Stokes parameters are obtained from ratios of aperture sums measured in different cameras and at different exposures, as in other polarimeters \citep[e.g.,][]{Kawabata+1999pasp, Watanabe+2012}. These ratios are expected to be only weakly affected by flat-fielding effects as long as the relative image positions are preserved. We further verified that polarization results obtained with and without flat-field correction show no significant difference, supporting our conclusion.



\subsection{Polarimetric performance}
\label{subsec:performance}
Finally, we evaluated the polarimetric performance of SQUIDPOL by observing polarimetric standard stars. Three types of standards were observed: a 100\% polarized reference light source, unpolarized standard stars (UPs), and strongly polarized standard stars (SPs). The observed targets and observation dates are listed in Table~\ref{tab:test_obs}. 
The polarimetric efficiency was measured by passing bright starlight through a WGF mounted in filter wheel~1 and treating the resulting light as a 100\% polarized light source. 

In the original design of SQUIDPOL, we assumed that the NPBS would divide the incident light into two beams without altering its polarization state, and the data analysis was conducted based on this assumption. However, during the instrument performance evaluation, we found that the commercial NPBS from Edmund Optics installed in SQUIDPOL introduces a measurable effect on the polarization state of the split light. To investigate this effect, we observed two SPs (HD 7927 and HD 25443) while rotating the HWP to four positions (0$^\circ$, 22.5$^\circ$, 45$^\circ$, and 67.5$^\circ$). We independently derived the polarization degree (\(P\)) and polarization angle (\(\theta_P\)) for the reflected beam from the NPBS (using images from Cameras~1 and 2) and for the transmitted beam (using images from Cameras~3 and 4). Observations were conducted in four filters ($B$-, $V$-, $R_\mathrm{C}$-, and $I_\mathrm{C}$-bands). We found that the polarization degrees \(P\) derived from the reflected and transmitted beams are mutually consistent, whereas the polarization angles \(\theta_P\) exhibit a systematic offset between the two beams. This offset remains constant regardless of the observed standard star. The measured \(\theta_P\) offsets are summarized in Table~\ref{tab:NPB_effect}. Such effects can be corrected during the data reduction process. We note that the observed offset in \(\theta_P\) between the reflected and transmitted beams should be taken into account when using similar commercial NPBS components in polarimetric instruments.

\begin{table}[htbp]
\centering
\caption{Polarization angles measured from the reflected (Camera~1\&2) and transmitted (Camera~3\&4) beams of the NPBS.}
\label{tab:NPB_effect}
\setlength{\tabcolsep}{2.3pt}
\begin{tabular}{lrrr}
\hline
Object$^{a}$& $\theta_\mathrm{P,\, Cam~1\&2}^{b}$ & $\theta_\mathrm{P,\, Cam~3\&4}^{c}$ & $\Delta \theta_\mathrm{P}^{d}$\\
&(deg)&(deg)&(deg)\\
\hline
100\% polarized & $-9.27 \pm 0.19$ & $1.16 \pm 1.05$& $10.43 \pm 1.06$\\
HD 25443& $-35.38\pm 0.82 $ & $46.78\pm 0.81$& $11.40\pm 1.16 $\\
HD 7927&$79.05 \pm 0.46$ & $91.82 \pm 2.10$&$12.77 \pm 2.15$\\
\hline
weighted mean&&&$11.09\pm 0.74$\\
\hline
\end{tabular}
\tabnote{
$^{a}$ Polarized light obtained using either the WGF or SPs.
$^{b}$ Measurement derived from Cameras~1 and 2.
$^{c}$ Measurement derived from Cameras~3 and 4.
$^{d}$ Difference between $\theta_\mathrm{P}$ measured from Cameras~1\&2 and Cameras~3\&4.
All measurements were obtained in the $R_\mathrm{C}$ band.
}
\end{table}

\begin{table}[h]
    \centering
\caption{Wavelength dependence of the $\theta_{\mathrm{P}}$ offset associated with the NPBS}
    \label{tab:npb_offset_summary_targets}
    \begin{tabular}{cr}
    \hline
    \textbf{Band} & $\theta_{\mathrm{offset}}$$^{\mathrm{a}}$\\
    \hline
    $B$ & $9.90 \pm 0.96$ \\
    $V$ & $10.36 \pm 0.36$ \\
    $R_\mathrm{C}$ & $11.09 \pm 0.74$ \\
    $I_\mathrm{C}$ & $11.07 \pm 0.60$ \\
    \hline
    \end{tabular}
    \tabnote{$^{a}$ Difference in the polarization position angle between the reflected and transmitted beams of the NPBS. The $R_\mathrm{C}$-band offset was determined using 100\% polarized light, HD~25443, and HD~7927, whereas the offsets in the other bands were determined using the 100\% polarized light only.}
\end{table}

After correcting the $\theta_{P}$ offset induced by the NPBS, we determined SQUIDPOL's polarization efficiency ($\mathrm{eff}$), instrumental polarization ($q_\mathrm{inst}$ and $u_\mathrm{inst}$), and polarization angle offset ($\theta_\mathrm{off}$). The methods for deriving these parameters are described in previous studies (e.g., \citealt{Ishiguro+2017AJ, Kawabata+1999pasp} ~and references therein). The derived calibration parameters of SQUIDPOL are summarized in Table \ref{tab:calibration}.

Figure \ref{fig:UP} shows the histogram of UPs observed by SQUIDPOL. The values are corrected for efficiency and the position angle offset induced by the NPB. As expected for unpolarized sources, the measured $q$ and $u$ values are distributed around zero in all bands. We fitted these histograms with 1D-Gaussian functions. The derived $1\sigma$ of $P$ ranges between 0.1\% and 0.15\% for all bands except B-band. The lower accuracy in the $B$-band is attributed to the employment of commercial optical components that are optimized for the $V$- and $R-$ bands. This shows that SQUIDPOL achieves a polarimetric precision comparable to our target precision of $\sim$ 0.2\%.

\begin{table}[h]
    \centering
    \caption{Calibration parameters of SQUIDPOL after correction of the NPBS-induced polarization-angle offset.}
    \label{tab:calibration}
    \small
    \setlength{\tabcolsep}{1.4pt}
    \begin{tabular}{lrrrr}
        \toprule
        Band   & $B$ & $V$ & $R_\mathrm{C}$ & $I_\mathrm{C}$ \\
        \midrule
        $p_\mathrm{eff,r}$ (\%) 
            & $70.86 \pm 0.95$
            & $60.94 \pm 1.25$
            & $98.48 \pm 0.57$
            & $98.68 \pm 0.68$\\
        $p_\mathrm{eff,t}$ (\%) 
            & $76.94 \pm 0.74$
            & $93.85 \pm 0.96$
            & $95.58 \pm 0.51$
            & $98.66 \pm 0.24$\\
        $q_\mathrm{inst}$ (\%) 
            & $-0.08 \pm 0.20$
            & $0.01 \pm 0.15$
            & $-0.04 \pm 0.15$
            & $0.00 \pm 0.09$\\
        $u_\mathrm{inst}$ (\%) 
            & $-0.06 \pm 0.23$
            & $0.01 \pm 0.14$
            & $0.08 \pm 0.15$
            & $-0.06 \pm 0.11$\\
        $\theta_\mathrm{off}$ ($^\circ$) 
            & $-3.87 \pm 1.68$
            & $ -1.17\pm 1.10$
            & $-0.73 \pm 1.35$
            & $-1.22 \pm 0.96$\\
        \bottomrule
    \end{tabular}
        \tabnote{Uncertainties represent 1$\sigma$ errors and have the same units as the corresponding quantities. $p_\mathrm{eff,r}$ and $p_\mathrm{eff,t}$ indicate the polarimetric efficiencies for the reflected and transmitted light components at the NPBS, respectively. }
\end{table}

\begin{figure}
\begin{subfigure}{\linewidth}
\label{fig:UP}
\includegraphics[width=\linewidth]{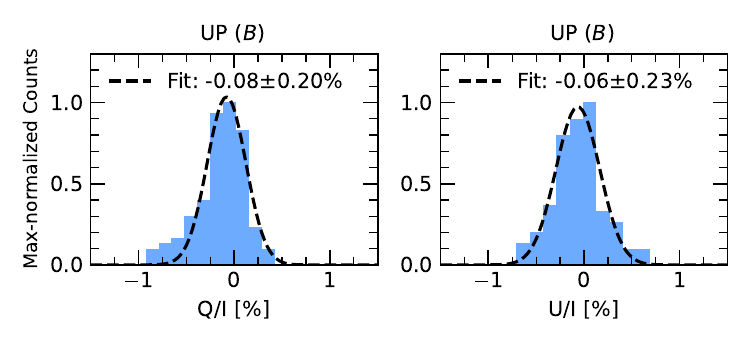}
\end{subfigure}

\vspace{-3mm}

\begin{subfigure}{\linewidth}
\includegraphics[width=\linewidth]{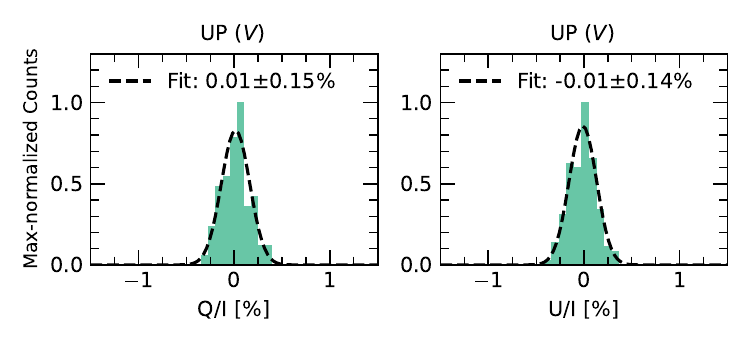}
\end{subfigure}

\vspace{-3mm}

\begin{subfigure}{\linewidth}
\includegraphics[width=\linewidth]{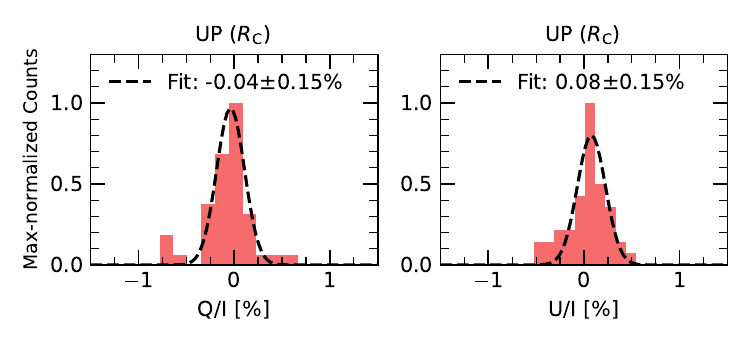}
\end{subfigure}

\vspace{-3mm}

\begin{subfigure}{\linewidth}
\includegraphics[width=\linewidth]{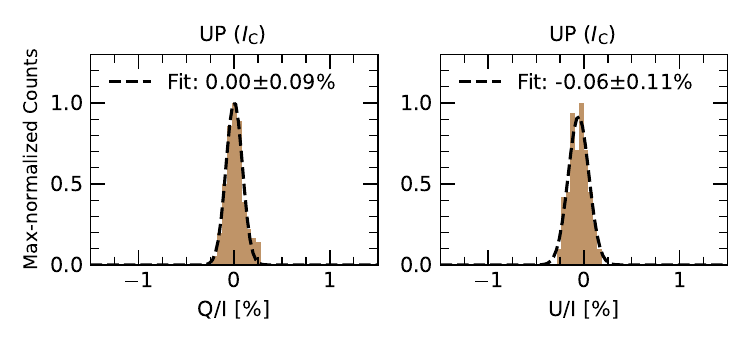}
\end{subfigure}

\caption{Histograms of $q$ and $u$ measured for unpolarized standards (Caph and $\gamma$~Bo\"otis) with SQUIDPOL between 2025 December 30 and 2026 Januaray 28 UT, after 3$\sigma$-clipping.
From top to bottom, the panels correspond to the $B$-, $V$-, $R_\mathrm{C}$-, and $I_\mathrm{C}$-bands. 
All values are shown after efficiency correction and before the instrument polarization correction, and the dashed lines indicate the fitting result using Gaussian function. Each histogram is normalized so that its maximum value is unity.}
\label{fig:UP}
\end{figure}

\begin{table*}[t]
    \centering
    \caption{Summary of key parameters of the telescope and SQUIDPOL}
    \label{tab:squidpol_param}
    \begin{tabular}{ll}
        \toprule
        \multicolumn{2}{c}{\textbf{The 60-cm telescope at Pyeongchang Observatory of SNU}} \\ 
        \midrule
        Telescope location 
            & $127^{\circ}03'00.0''$ E, $37^{\circ}34'59.9''$ N \\
        Telescope type 
            & Ritchey--Chr\'etien telescope (equatorial mount)\\
        Primary mirror diameter
            & 0.6 m \\
        Focal length 
            & 4200 mm ($f/7$) \\
        \midrule
        \multicolumn{2}{c}{\textbf{SQUIDPOL}} \\ 
        \midrule
        Camera 
            & ASI 294MM mono (2072 $\times$ 1411$^{a}$) \\
        Pixel size$^{b}$ 
            & 9.2 $\mu$m \\
        Pixel scale$^{c}$
            & 0.453 $''$ pixel$^{-1}$ \\
        FOV per camera
            & $15.6 \times 10.6$ arcmin$^{2}$ \\
        Effective FOV $^{d}$
            & 110.6 arcmin$^{2}$ ($\sim$ $13 \times 8.5$ arcmin$^{2}$) \\
        Filter wheel configuration
            & (Wheel~1) $B$, $V$, $R_\mathrm{C}$, $I_\mathrm{C}$; \\
            & (Wheel~2) WGF, light-blocking plate, $U^{e}$  \\
        \bottomrule
    \end{tabular}
    \tabnote{
    $^{a}$ Resolution in the 4$\times$4 binning mode (originally 8288 $\times$ 5644).
    $^{b}$ Effective pixel size in the 4$\times$4 binning mode.
    $^{c}$ Effective pixel scale in the 4$\times$4 binning mode.
    $^{d}$ Effective field of view observed by all four cameras.
    $^{e}$ The $U$-band filter is available only for photometric observations.
    }
\end{table*}

\section{Summary\label{sec:summary}}

We have developed SQUIDPOL, a low-cost imaging polarimeter based on commercially available optical components, designed to perform simultaneous optical polarimetry using four CMOS imagers. Each imager records one of four polarization states, enabling efficient and stable measurements without mechanical modulation during exposures. SQUIDPOL has been installed on the 60-cm telescope at the Pyeongchang campus and has been operating successfully in regular observations.

We conducted a comprehensive in-situ performance evaluation of the instrument, including measurements of polarization efficiency, instrumental polarization, and polarization angle calibration. The measured performance parameters are generally consistent with expectations from optical simulations, demonstrating that SQUIDPOL is well suited for scientific polarimetric observations. During this evaluation, we also identified a systematic polarization angle offset introduced by the commercial NPBS, which can be reliably corrected during data reduction.

SQUIDPOL is now being used for survey observations of small Solar System bodies, including asteroids, comets, and planetary satellites, with the aim of statistically investigating their surface and dust properties through polarimetry. In addition to its scientific applications, the instrument serves as a valuable platform for education and hands-on training in observational astronomy. Future upgrades are planned to further improve its performance and to expand its observational capabilities.



\acknowledgments
We would like to express our sincere gratitude to all those involved in the establishment of the Pyeongchang Observatory of Seoul National University (SNU) and the initiation of its observational activities. In particular, we thank the faculty and staff members of the SNU Astronomy Program and SNU Astronomy Research Center (SNUARC) for their understanding, encouragement, and continued support for the development of SQUIDPOL. The development of SQUIDPOL and associated activities were supported by a grant from the Korean National Research Foundation (NRF), funded by the Korean government (MEST; No. 2023R1A2C1006180). During the installation and observation period, we benefited from the generous logistical support provided by the staff of the Korean VLBI Network (KVN) Pyeongchang Station. J. Geem acknowledges support from the Carl Trygger Foundation (CTS 24:3838).


\appendix


\bibliography{SQ}

\end{document}